\documentclass[letterpaper, journal]{IEEEtran}
\usepackage{amsmath, amssymb}
\usepackage{url, doi}
\usepackage{algorithm, algorithmic}
\usepackage{graphicx, xcolor, subcaption, float}
\usepackage{harpoon}
\usepackage{hyphenat}
\usepackage{tikz, pgfplots}
\pgfplotsset{compat=1.17}

\def\x{{\mathbf x}}

\mathchardef\mhyphen="2D
\def\cpar{\hss\egroup\line\bgroup\hss}

\newcommand{\mb}[1]{\mathbf{#1}}

\DeclareMathOperator*{\argmin}{arg\,min}

\newcommand{\fig}[1]{Fig.~\ref{fig:#1}}

\title{Total Variation Regularization for Tomographic Reconstruction of Cylindrically Symmetric Objects}

\author{
Maliha Hossain, Charles A. Bouman, Brendt Wohlberg
  \thanks{
    Maliha Hossein and Charles A. Bouman are with the School of ECE and BME of Purdue University, West Lafayette, IN 47907 USA. E-mail: \texttt{mhossain@purdue.edu}, \texttt{bouman@purdue.edu}
  }
  \thanks{
    Brendt Wohlberg is with Theoretical Division, Los Alamos National Laboratory, Los Alamos, NM 87545 USA. E-mail: \texttt{brendt@lanl.gov}
    }
  \thanks{
    This research was supported by the Laboratory Directed Research and Development program of Los Alamos National Laboratory under project numbers 20200061DR and 20230771DI.
  }
}

\begin{document}

\maketitle

\begin{abstract}
Flash X-ray computed tomography (CT) is an important imaging modality for characterization of high-speed dynamic events, such as Kolsky bar impact experiments for the study of  mechanical properties of materials subjected
to impulsive forces. Due to experimental constraints, the number of X-ray views that can be obtained is typically very sparse in both space and time, requiring strong priors in order to enable a CT reconstruction.  In this paper, we propose an effective method for exploiting the cylindrical symmetry inherent in the experiment via a variant of total variation (TV) regularization that operates in cylindrical coordinates, and demonstrate that it outperforms competing  approaches.
\end{abstract}

\begin{IEEEkeywords}
Sparse View Tomography,
Total Variation Regularization
Cylindrical Symmetry
\end{IEEEkeywords}

\section{Introduction}
\label{sec:intro}

\begin{figure}[htb]
\begin{minipage}[b]{.90\linewidth}
  \centering
  \centerline{\includegraphics[width=5.70cm]{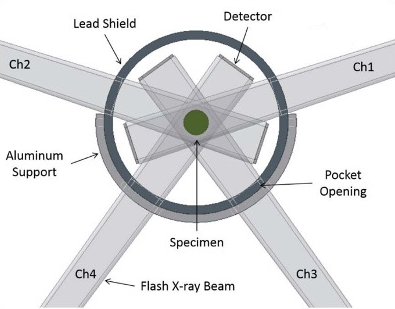}}
  \vspace{-0.1cm}
\end{minipage}
\caption{
Typical configuration of a Kolsky bar experiment.}
\label{fig:kolsky}
\end{figure}

Accurate characterization of material behavior is critical for ensuring
structural integrity in automotive, biomedical, and civil engineering applications, where specimens are subjected to high stress and strain rates.
The Kolsky bar experiment~\cite{liao2018flash} is among
the most common methods for studying material properties under high stress.
During these experiments, cylindrical specimens are mounted between clamps
and subjected to brief impulsive forces over a span of less than a millisecond. The
resulting fracture formation in the specimens is characterized via computed tomography (CT) reconstruction of flash X-ray measurements. The dynamic nature of these experiments necessitates in situ X-ray source-detector pairs, as illustrated in~\fig{kolsky}, making ultra sparse view sampling unavoidable, so that  regularized CT reconstruction, such as model based iterative reconstruction (MBIR)~\cite{bouman1996unified}, is essential to address the highly ill-posed nature of the associated inverse problem.

The choice of an appropriate regularizer (prior model) is essential to obtaining good performance in such cases. Priors based on relatively simple models of local regularity, such as total variation (TV)~\cite{rudin1992nonlinear} and the q-generalized Gaussian Markov random field (qGGMRF)~\cite{thibault2007three} are more than two decades old, and are still widely used today, although their performance has been eclipsed by more recent methods for complex classes of imagery such as natural images. Prominent examples of such methods include plug-and-play priors (PPP)~\cite{venkatakrishnan2013plug, kamilov-2023-plug} and consensus equlibrium (CE)~\cite{buzzard2018plug} together with denoisers such as BM3D~\cite{dabov2007image} and BM4D~\cite{maggioni2012nonlocal}. None of these methods, however, are able to directly exploit the cylindrical symmetry (which is at least approximately retained after impact) of the sample in the Kolsky bar experiment. A recent approach developed specifically for this type of problem employed a CE framework with BM3D denoisers combined via multi-slice fusion~\cite{majee20194d}, together with a rotation-and-average operator to promote the expected circular symmetry in the plane perpendicular to the axis of the cylinder~\cite{hossain2020ultrasparse}. While it outperformed competing methods, this method is computationally expensive, and the approach to promoting circular symmetry is somewhat \emph{ad hoc}.
The goal of the work reported here was the development of a less \emph{ad hoc} method for exploiting the cylindrical symmetry inherent in the problem. Since the specimens in Kolsky bar experiments are typically
3D printed shapes with circular cross sections, and  are homogeneous on the imaging scale, TV in cylindrical coordinates is a natural choice of regularizer.

To the best of our knowledge, this is the first work to propose a variant of TV for regularization of cylindrically symmetric structures. The most closely related TV variants of which we are aware are those that are designed to be directionally adaptive to local image structure~\cite{bayram2012directional, zhang2013adaptive, zhang2013edge, ehrhardt2016multicontrast}, some of which are also based on the concept of projection of the gradient estimates onto a locally varying coordinate system.
These methods are not very well suited to our application because the adaptivity depends on an estimate of dominant local image orientation, which is not straightforward to integrate into the CT reconstruction, and does not exploit the prior knowledge of cylindrical symmetry.

With respect to the specific application to CT reconstruction, while~\cite{asaki-2005-abel} superficially appears to address the same problem, it is, in a sense, the opposite problem: the reconstruction has a cylindrical discretization, and the TV term is radially weighted so that its integral form is equivalent to an integral in a Cartesian coordinate system. A number of other authors have considered variants of TV with directional preference, e.g.~\cite{chen-2013-limitedangle, guo2017image, sheng2020sequential}, for suppression of limited-angle CT reconstruction artifacts, but none of these involve any form of symmetry, and the preferential directions are all defined globally rather than locally. A transformation into polar coordinates is used in~\cite{sijbers2004reduction} for correction of micro-CT ring artifacts, but the correction is performed directly via median filtering rather than via regularization methods.

\section{Cylindrical Total Variation}
\label{sec:ctv}

\begin{figure*}[htb]
\centering
\subcaptionbox{Polar\label{fig:pcpcnva}}
{\includegraphics[height=0.3\linewidth]{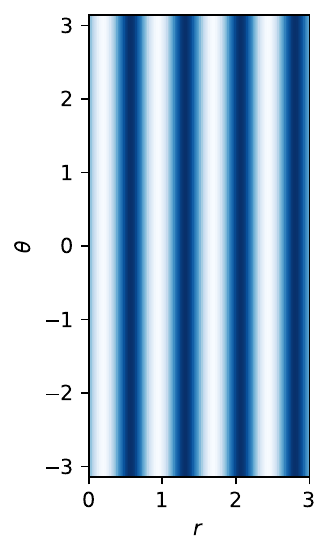}}
\subcaptionbox{Cartesian\label{fig:pcpcnvb}}
{\includegraphics[height=0.3\linewidth]{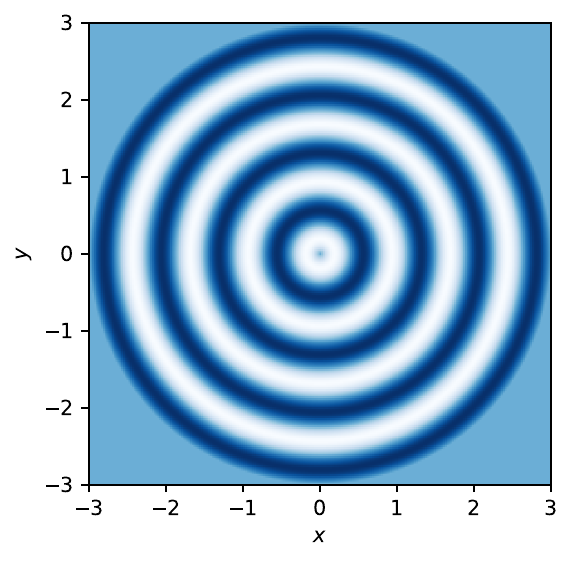}}
\subcaptionbox{Coordinate conversion error\label{fig:pcpcnvc}}
{\includegraphics[height=0.3\linewidth]{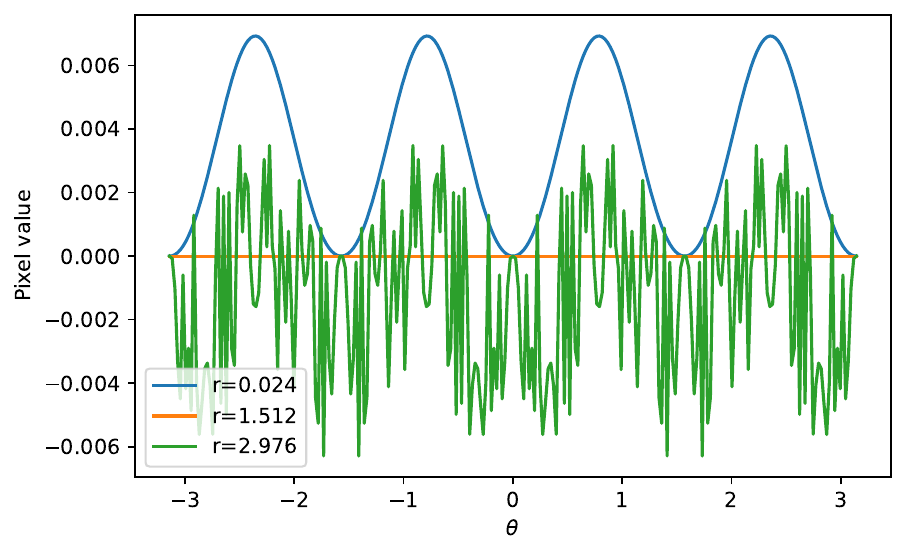}}
\caption{Errors resulting from interpolation required for conversion between polar and Cartesian coordinates. Image (\subref{fig:pcpcnva}) is defined in polar coordinates, on a $255 \times 128$ grid, as being constant along the angular axis and sinusoidal along the radial axis, while image (\subref{fig:pcpcnvb}) is the result of a conversion to Cartesian coordinates on a $255 \times 255$ grid. The plots in (\subref{fig:pcpcnvc}) indicate the difference between the image in \subref{fig:pcpcnva} and the image in \subref{fig:pcpcnvb} after conversion back to polar coordinates. The conversion is generally of acceptable accuracy, but significant errors for radii near the origin and maximum of the range are very difficult to avoid. Third degree splines were used for the coordinate conversions in this example; significantly larger errors
are observed for linear interpolation.
}
\label{fig:coordtrans}
\end{figure*}

\usetikzlibrary{math}
\tikzmath{
\r0 = 1.5;
\thet0 = 25;
\r1 = 2.5;
\r2 = 2.0;
\thet2 = \thet0 + 20;
\x0 = \r0 * cos(\thet0);
\y0 = \r0 * sin(\thet0);
\xc0 = \x0;
\yc0 = \y0 + \r1;
\xc1 = \x0 + \r1;
\yc1 = \y0;
\thet1 = \thet0 + 90;
\xp0 = \x0 + \r1 * cos(\thet1);
\yp0 = \y0 + \r1 * sin(\thet1);
\xp1 = \x0 + \r1 * cos(\thet0);
\yp1 = \y0 + \r1 * sin(\thet0);
\xvc = \x0 + \r2 * cos(\thet2);
\yvc = \y0 + \r2 * sin(\thet2);
\xvpr = \x0 + \r2 * cos(\thet2 - \thet0) * cos(\thet0);
\yvpr = \y0 + \r2 * cos(\thet2 - \thet0) * sin(\thet0);
\xvpp = \x0 + \r2 * sin(\thet2 - \thet0) * cos(\thet1);
\yvpp = \y0 + \r2 * sin(\thet2 - \thet0) * sin(\thet1);
}

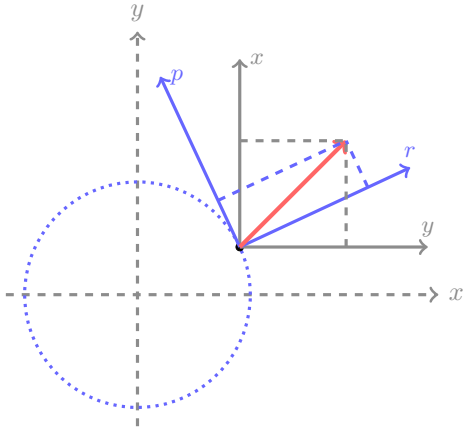
\begin{figure}[htb]
\centering
\begin{tikzpicture}[scale=1]
\draw[->, very thick, dashed, color=gray!90] (-1.75,0)--(4,0) node[right]{$x$};
\draw[->, very thick, dashed, color=gray!90] (0,-1.75)--(0,3.5) node[above]{$y$};

\draw[dotted, color=blue!60, very thick](0,0) circle (\r0);

\filldraw[color=black, fill=black](\x0, \y0) circle (0.05);

\draw[->, color=gray!90, very thick] (\x0, \y0)--(\xc0, \yc0) node[right]{\small $x$};
\draw[->, color=gray!90, very thick] (\x0, \y0)--(\xc1, \yc1) node[above]{\small $y$};

\draw[->, color=blue!60, very thick] (\x0, \y0)--(\xp0, \yp0) node[right]{\small $p$};
\draw[->, color=blue!60, very thick] (\x0, \y0)--(\xp1, \yp1) node[above]{\small $r$};

\draw[->, color=red!60, ultra thick] (\x0, \y0)--(\xvc, \yvc);

\draw[color=gray!90, dashed, very thick] (\xvc, \y0)--(\xvc, \yvc);
\draw[color=gray!90, dashed, very thick] (\x0, \yvc)--(\xvc, \yvc);

\draw[color=blue!60, dashed, very thick] (\xvpr, \yvpr)--(\xvc, \yvc);
\draw[color=blue!60, dashed, very thick] (\xvpp, \yvpp)--(\xvc, \yvc);
\end{tikzpicture}
\caption{
Gradient projection from global Cartesian to local polar coordinate system. The Cartesian coordinate axes are denoted by $x$ and $y$, and the polar coordinate axes are denoted by $r$ and $p$. The red vector and the corresponding grey dashed lines represent an example of a difference vector computed within the Cartesian coordinate system. The corresponding blue dashed lines indicate the representation obtained by projecting the difference vector onto the local polar coordinate system.}
\label{fig:grad_project}
\end{figure}

The most obvious approach to constructing TV regularizer in cylindrical coordinates is to apply the standard TV regularization function to a volume that has been transformed into cylindrical coordinates. Unfortunately, however, such an approach is problematic, due to the artifacts resulting from transforming a sampled representation in Cartesian (rectangular) coordinates to a corresponding sampled representation in polar coordinates. These errors are a result of the interpolation required for the coordinate conversion, as illustrated in~\fig{coordtrans}.

A much more effective approach is possible, however, by observing that the gradients in cylindrical coordinates can be computed \emph{after} estimation of the gradients in Cartesian coordinates, thereby avoiding any interpolation, as illustrated in~\fig{grad_project} for the slightly simpler case of polar coordinates. This conversion to cylindrical coordinates can be expressed as follows. Consider a point in the sampled volume with angle $\theta$ (with respect to a user-selected center of the polar coordinate system) in the $x\!-\!y$ plane and with local gradient $\mb{\Delta}_{\text{rec}} = (\Delta_x, \Delta_y, \Delta_z)$ in recangular coordinates. If we define local unit-norm coordinate axes in cylindrical coordinates as
\begin{equation}
    \mb{p} = \left( \! \begin{array}{r}
        -\cos(\theta) \\ \sin(\theta) \\ 0
    \end{array} \! \right)
    \quad
    \mb{r} = \left( \!\begin{array}{r}
        \sin(\theta) \\ \cos(\theta) \\ 0
    \end{array} \! \right)
    \quad
    \mb{z} = \left( \! \begin{array}{r}
        0 \\ 0 \\ 1
    \end{array} \! \right) \,,
\end{equation}
then the components of the gradient in the cylindrical coordinates are
\begin{equation}
    \mb{\Delta}_{\text{cyl}} = (\Delta_p, \Delta_r, \Delta_z) = (\mb{p}^T \mb{\Delta}_{\text{rec}}, \mb{r}^T \mb{\Delta}_{\text{rec}}, \mb{z}^T \mb{\Delta}_{\text{rec}}) \;.
\end{equation}

The total variation regularization function is then computed as usual, but using the gradient in cylindrical rather than rectangular coordinates. It must be emphasized that while isotropic total variation is usually preferred over the anisotropic form (see e.g.~\cite{lou-2015-weighted} for a discussion of the difference), it should be avoided here: since isotropic TV computes local gradient magnitudes prior to summing over all samples in the array, and since the local coordinate transform is a magnitude-preserving rotation, the isotropic total variation in rectangular and cylindrical coordinates are identical. The proposed cylindrical TV functional is therefore
\begin{equation}
    \text{CTV}(\mb{x}) = \lambda_p \| C_p \mb{x} \|_1 + \lambda_r \| C_r \mb{x} \|_1  + \lambda_z \| C_z \mb{x} \|_1 \;,
\end{equation}
where the operators $C_p, C_r$, and $C_z$ compute the gradients in the angular, radial, and axial directions respectively. As in the case of standard TV, the local gradients should be computed using first-order (two-point) differences rather than higher-order gradient approximations, which can lead to undesirable artifacts.

\section{Ultra-Sparse View CT}
\label{sec:usvct}

We solve the CT problem via the regularized inverse problem
\begin{align}
\argmin_{\mb{x}} \; (1/2) \, \| A \mb{x} - \mb{y} \|_2^2 + \text{CTV}(\mb{x}) \;,
\end{align}
where $A$ represents the CT projection operator and $\mb{y}$ is a vector representation of the sinogram consisting of four views.
This problem is solved using the Primal-Dual Hybrid Gradient (PDHG) algorithm~\cite{esser2010general} rather than the
Alternating Direction Method of Multipliers (ADMM) \cite{boyd2011distributed} since PHDG avoids the need for computationally-expensive conjugate gradient (CG) iterations to solve a large linear subproblem involving the CT projection operator and the $C_p$, $C_r$, and $C_z$ operators. In the following, we will refer to this approach as MBIR/CTV.

The performance of the proposed method was compared with the following alternative approaches:
\begin{LaTeXdescription}
    \item[MBIR/TV] The problem is solved via standard MBIR formulation (i.e. regularized inversion) with a standard (recctangular coordinate) TV prior.
    \item[PPP/BM3D] The problem is solved via a plug-and-play priors (PPP) framework~\cite{sreehari-2016-plug} with a BM3D image denoiser~\cite{dabov2007image} applied to all slices in the $x\!-\!y$ plane.
    \item[PPP/BM4D] The problem is solved via a plug-and-play priors (PPP) framework~\cite{sreehari-2016-plug} with a BM4D volume denoiser~\cite{maggioni2012nonlocal} prior.
    \item[CE/MSF/BM3D] The problem is solved via a consensus equlibrium (CE) framework~\cite{sreehari-2016-plug} with a three separate BM3D image denoisers along $x\!-y\!$, $y\!-\!z$,  and $z\!-\!x$ planes integrated using the Multi-Slice Fusion (MSF)~\cite{majee20194d} technique. As opposed to PPP/BM3D,
    the effective prior model enforces regularity in the
    axial direction.
    \item[CE/MSF/BM3D/RA] Similar to the CE/MSF/BM3D approach described above, with an additional
    denoising operation performed by computing the
    weighted average of copies of the image rotated over a small range of angles in the $x\!-\!y$ plane to impose cylindrical symmetry~\cite{hossain2020ultrasparse}.
\end{LaTeXdescription}

All computational experiments were performed in Python, using the CT projection operator and MBIR/qGGMRF solver from the SVMBIR~\cite{svmbir-2020} package, the BM3D implementation~\cite{bm3d-2022} released with~\cite{makinen-2019-exact}, the BM4D implementation~\cite{bm4d-2022} released by the authors of~\cite{maggioni2012nonlocal}, and optimizers and TV regularizers from the the SCICO~\cite{balke-2022-scico} package.

\section{Results}
\label{sec:results}

In this section we present reconstructions from
simulated and experimental measurements for the flash X-ray scanner
in~\fig{kolsky} with 4 parallel beam views at
$18^{\circ}$, $162^{\circ}$, $234^{\circ}$, and $306^{\circ}$,
0.097 mm voxel spacing and 0.049 mm detector spacing.

We compare reconstructions made using MBIR/CTV against MBIR/TV
and the other priors outlined in the previous section
All the aforementioned reconstruction methods were initialized with
an MBIR$+$qGGMRF reconstruction, and implemented in Python
as extensions to the SCICO package~\cite{balke-2022-scico} on
Nvidia Tesla V100 GPU (32GB GPU memory) and 64GB system memory over 16 cores.
We also present results produced using MSFRI priors as in \cite{hossain2020ultrasparse}.

\begin{table}[htb]
\centering
\begin{tabular}{|l|r|c|r|r|}
\hline
Prior    & Iters   & \begin{tabular}[c]{@{}c@{}}PSNR\\ (dB)\end{tabular} &\begin{tabular}[c]{@{}c@{}}CPU time\\ (min)\end{tabular} & \begin{tabular}[c]{@{}c@{}}Wall time\\ (min)\end{tabular} \\ \hline
PPP/BM3D         & 40      & 13.10 & 6.13      & 5.22      \\ \hline
PPP/BM4D         & 40      & 13.28 & 57.53     & 6.13      \\ \hline
CE/MSF/BM3D     & 40      & 13.12 & 135.08    & 135.01    \\ \hline
CE/MSF/BM3D/RA  & 50      & 12.95 & 168.10    & 167.98    \\ \hline
MBIR/TV          & 500     & 12.93 & 5.00      & 1.32      \\ \hline
MBIR/CTV         & 500     & 13.49 & 6.22      & 1.73      \\ \hline
\end{tabular}
\caption{PSNR scores and run times for simulated results.}
\label{table:sim_table}
\end{table}

\begin{figure*}[htbp]
\begin{minipage}[b]{.245\linewidth}
  \centering
  \centerline{\includegraphics[height=3.60cm]{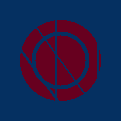}}
  \vspace{-0.1cm}
  \centerline{(a) Ground Truth}\medskip
\end{minipage}
\begin{minipage}[b]{.245\linewidth}
  \centering
  \centerline{\includegraphics[height=3.60cm]{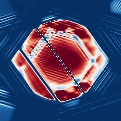}}
  \vspace{-0.1cm}
  \centerline{(c) PPP/BM3D}\medskip
\end{minipage}
\begin{minipage}[b]{0.245\linewidth}
  \centering
  \centerline{\includegraphics[height=3.60cm]{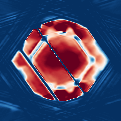}}
  \vspace{-0.1cm}
  \centerline{(e) PPP/BM4D}\medskip
\end{minipage}
\begin{minipage}[b]{.245\linewidth}
  \centering
  \centerline{\includegraphics[height=3.60cm]{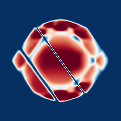}}
  \vspace{-0.1cm}
  \centerline{(d) CE/MSF/BM3D}\medskip
\end{minipage}
\begin{minipage}[b]{0.245\linewidth}
  \centering
  \centerline{\includegraphics[height=3.60cm]{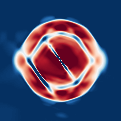}}
  \vspace{-0.1cm}
  \centerline{(g) CE/MSF/BM3D/RA}\medskip
\end{minipage}
\begin{minipage}[b]{0.245\linewidth}
  \centering
  \centerline{\includegraphics[height=3.60cm]{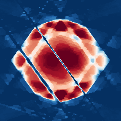}}
  \vspace{-0.1cm}
  \centerline{(f) MBIR/TV}\medskip
\end{minipage}
\begin{minipage}[b]{0.245\linewidth}
  \centering
  \centerline{\includegraphics[height=3.60cm]{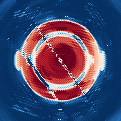}}
  \vspace{-0.1cm}
  \centerline{(h) MBIR/CTV}\medskip
\end{minipage}
\begin{minipage}[b]{0.245\linewidth}
  \hspace{5mm}\includegraphics[angle=90, origin=l, height=4.40cm]{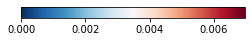}
  \vspace{-0.1cm}
\end{minipage}
\vspace{-0.2cm}
\caption{
Simulated results. (a) Cross section of ground truth for comparison with CT reconstructions (b) through (h) with various methods.}
\label{fig:sim_results}
\end{figure*}

A simulated phantom of a 3D column with radial, transverse and
concentric cracks is shown in~\fig{sim_results}.
The $121\times 121\times 100$ voxel volume was forward projected
and noise is added with standard deviation at 2$\%$ of the sinogram range.
Hyperparameters for each method were optimized using Ray Tune\footnote{See \url{https://docs.ray.io/en/latest/tune/index.html}.}

\begin{table}[htb]
\centering
\begin{tabular}{|l|r|r|r|}
\hline
Prior      & Iterations & \begin{tabular}[c]{@{}c@{}}CPU time\\ (min)\end{tabular} & \begin{tabular}[c]{@{}c@{}}Wall time\\ (min)\end{tabular} \\ \hline
PPP/BM3D         & 50      & 9.65      & 4.65  \\ \hline
PPP/BM4D         & 60      & 105.77    & 26.02 \\ \hline
CE/MSF/BM3D     & 30      & 98.45     & 95.27 \\ \hline
CE/MSF/BM3D/RA  & 50      & 154.45    & 89.74 \\ \hline
MBIR/TV          & 500     & 28.74     & 3.55  \\ \hline
MBIR/CTV         & 500     & 46.50     & 4.80  \\ \hline
\end{tabular}
\caption{Run times for experimental results.}
\label{table:exp_table}
\end{table}

\begin{figure*}[htbp]
\begin{minipage}[b]{0.30\linewidth}
  \centering
  \centerline{\includegraphics[height=5.7cm]{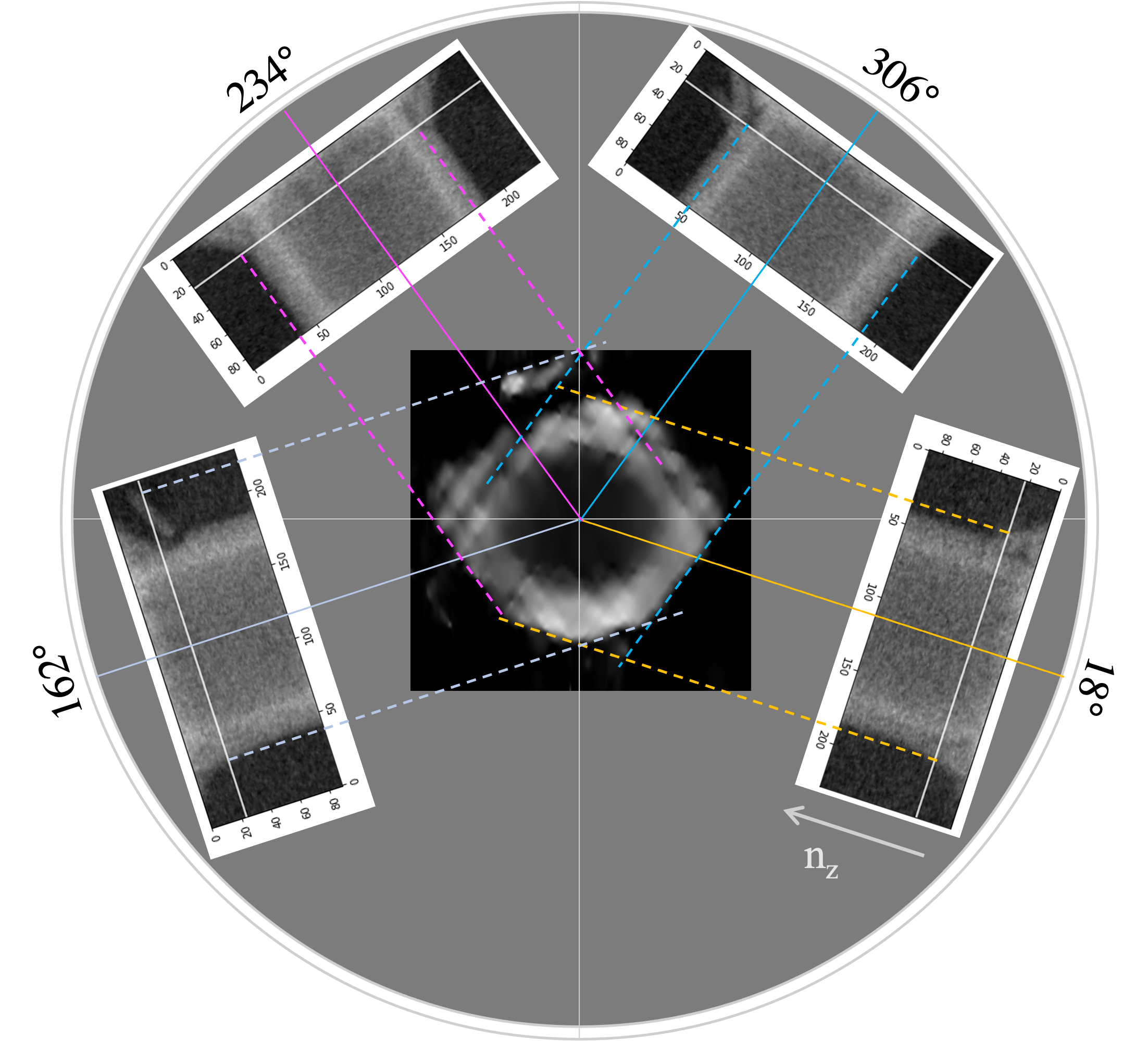}}
  \vspace{-0.1cm}
  \centerline{(b) PPP/BM3D}\medskip
\end{minipage}
\hfill
\begin{minipage}[b]{.30\linewidth}
  \centering
  \centerline{\includegraphics[height=5.7cm]{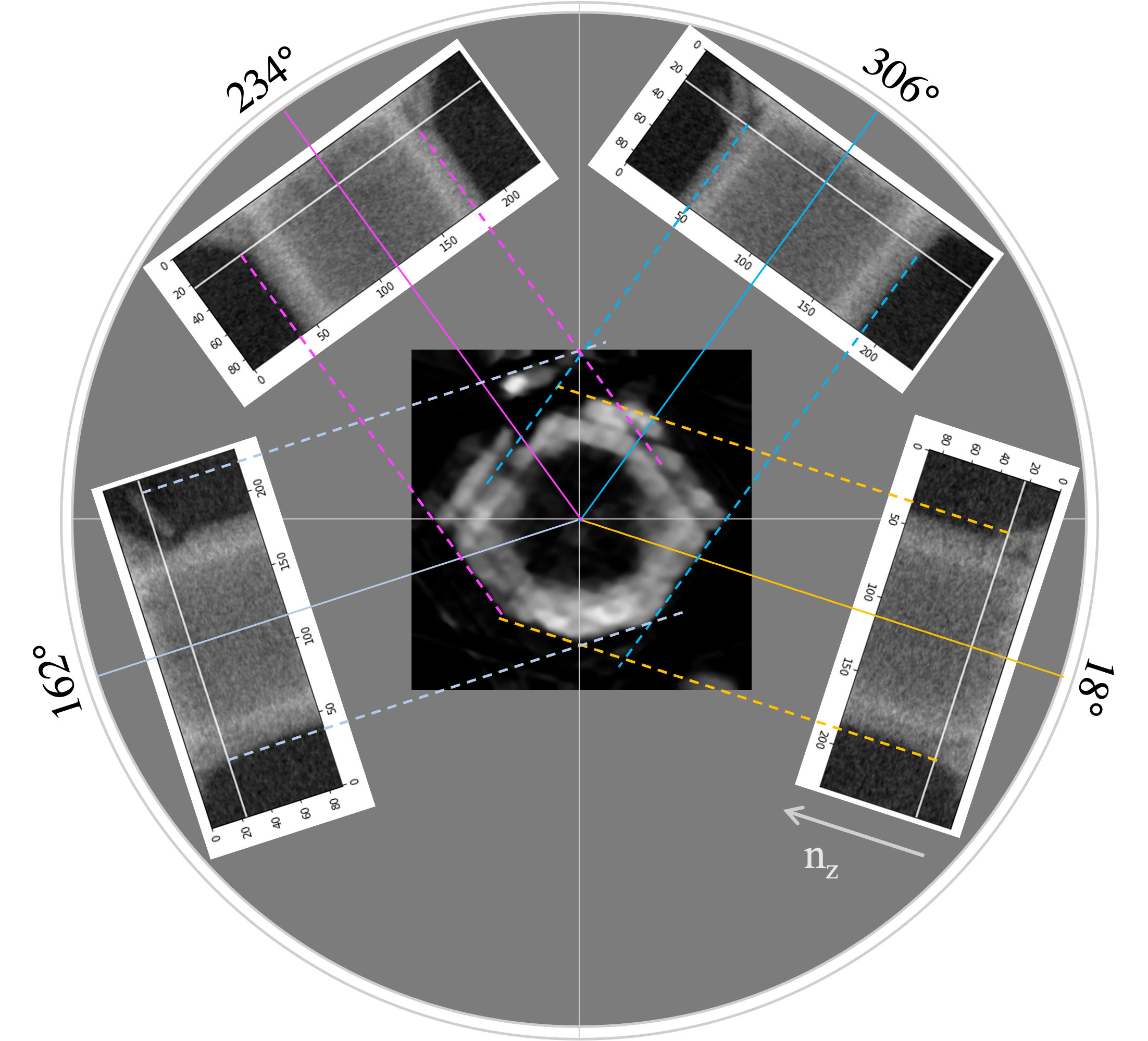}}
  \vspace{-0.1cm}
  \centerline{(c) PPP/BM4D}\medskip
\end{minipage}
\hfill
\begin{minipage}[b]{.30\linewidth}
  \centering
  \centerline{\includegraphics[height=5.7cm]{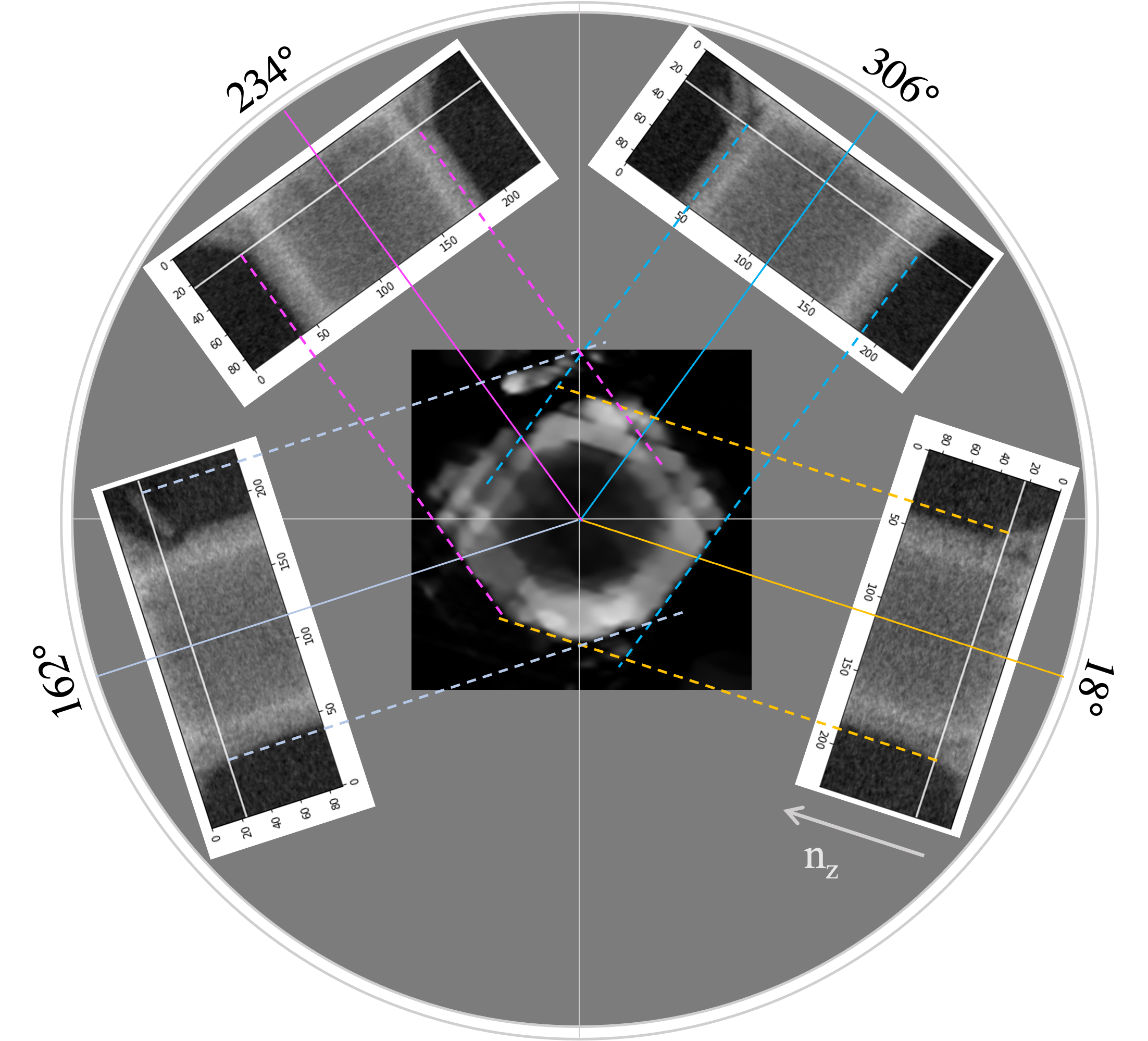}}
  \vspace{-0.1cm}
  \centerline{(d) CE/MSF/BM3D}\medskip
\end{minipage}
\hfill
\begin{minipage}[b]{.30\linewidth}
  \centering
  \centerline{\includegraphics[height=5.7cm]{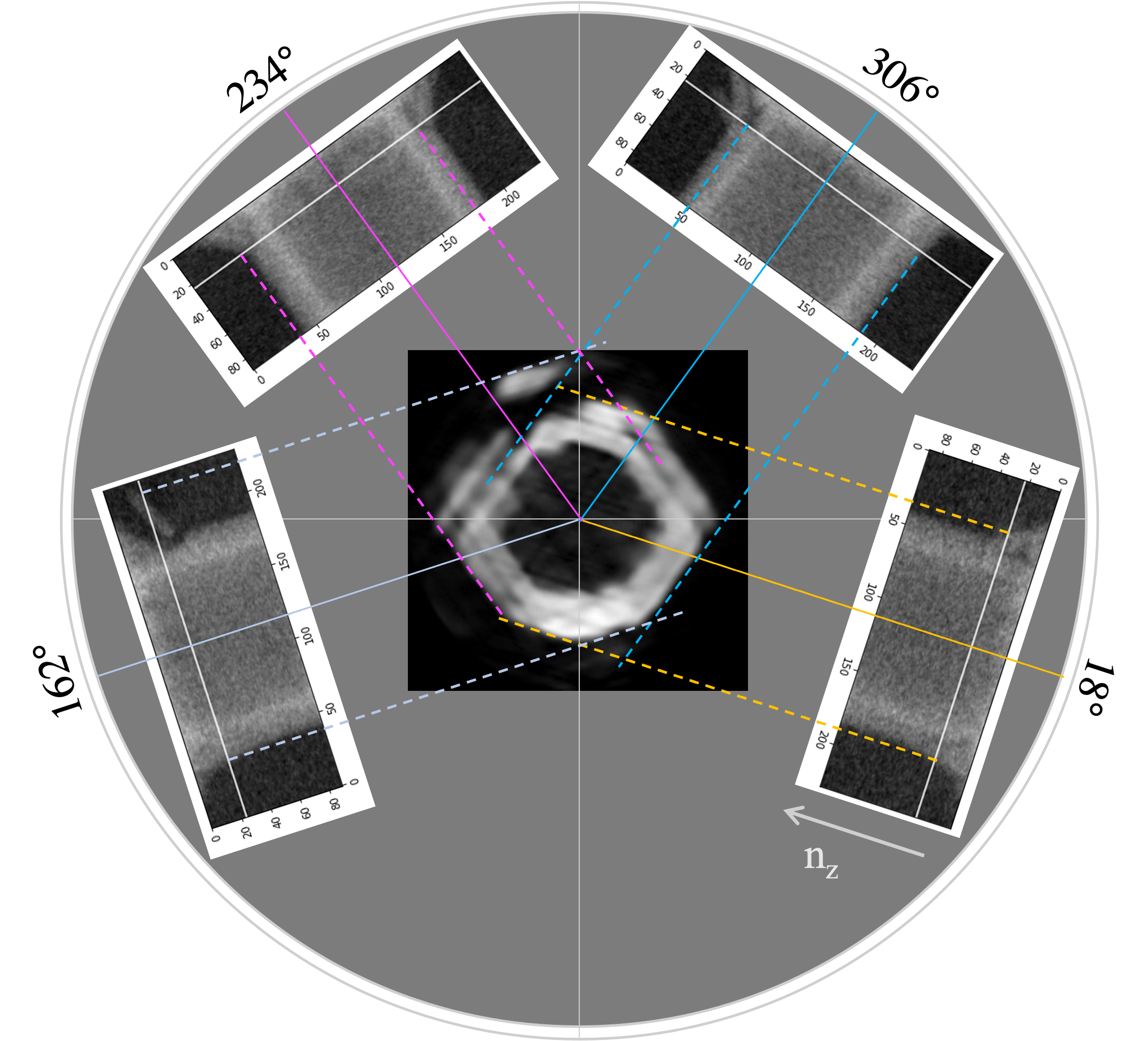}}
  \vspace{-0.1cm}
  \centerline{(e) CE/MSF/BM3D/RA}\medskip
\end{minipage}
\hfill
\begin{minipage}[b]{0.30\linewidth}
  \centering
  \centerline{\includegraphics[height=5.7cm]{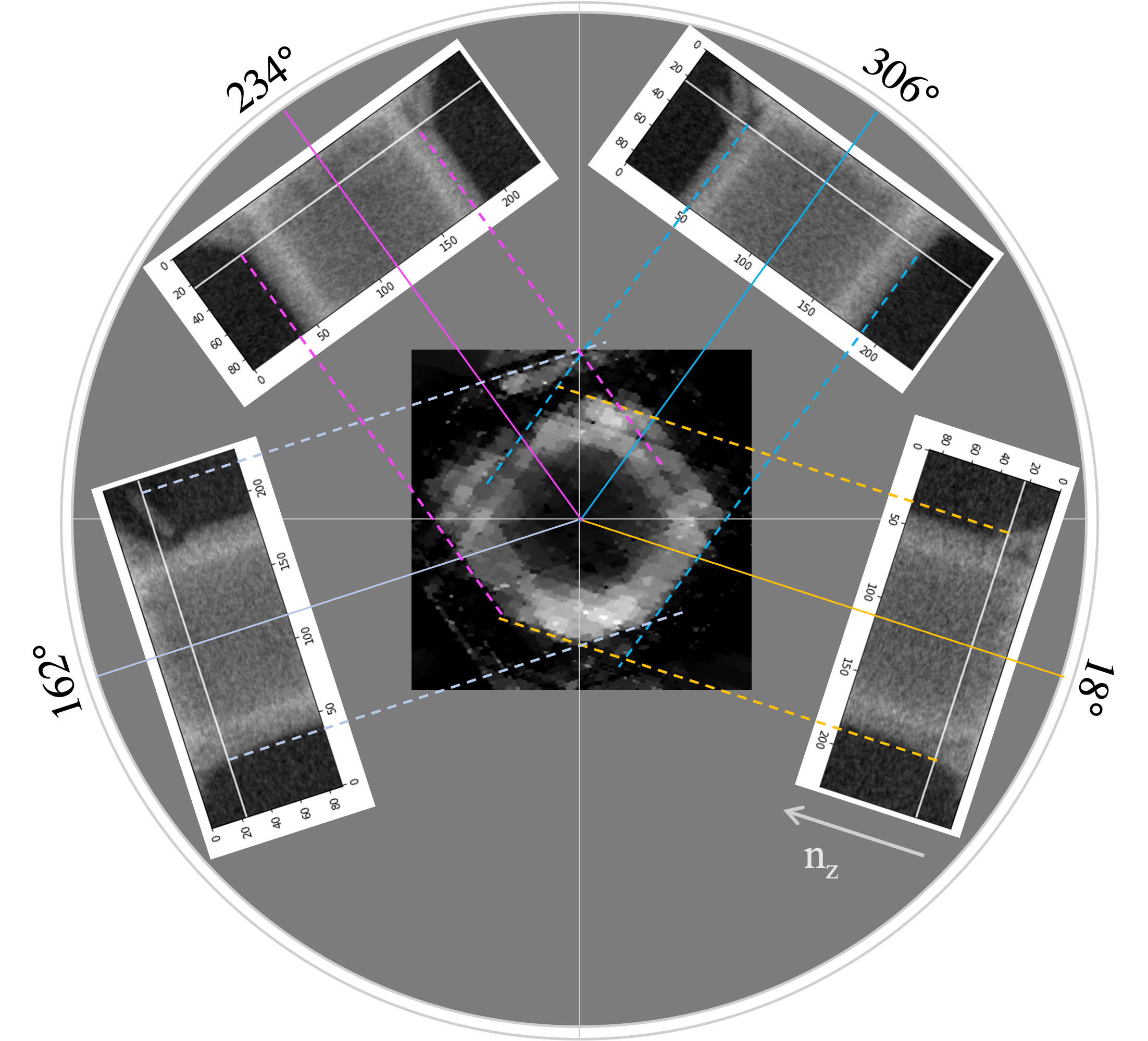}}
  \vspace{-0.1cm}
  \centerline{(f) MBIR/TV}\medskip
\end{minipage}
\hfill
\begin{minipage}[b]{.30\linewidth}
  \centering
  \centerline{\includegraphics[height=5.7cm]{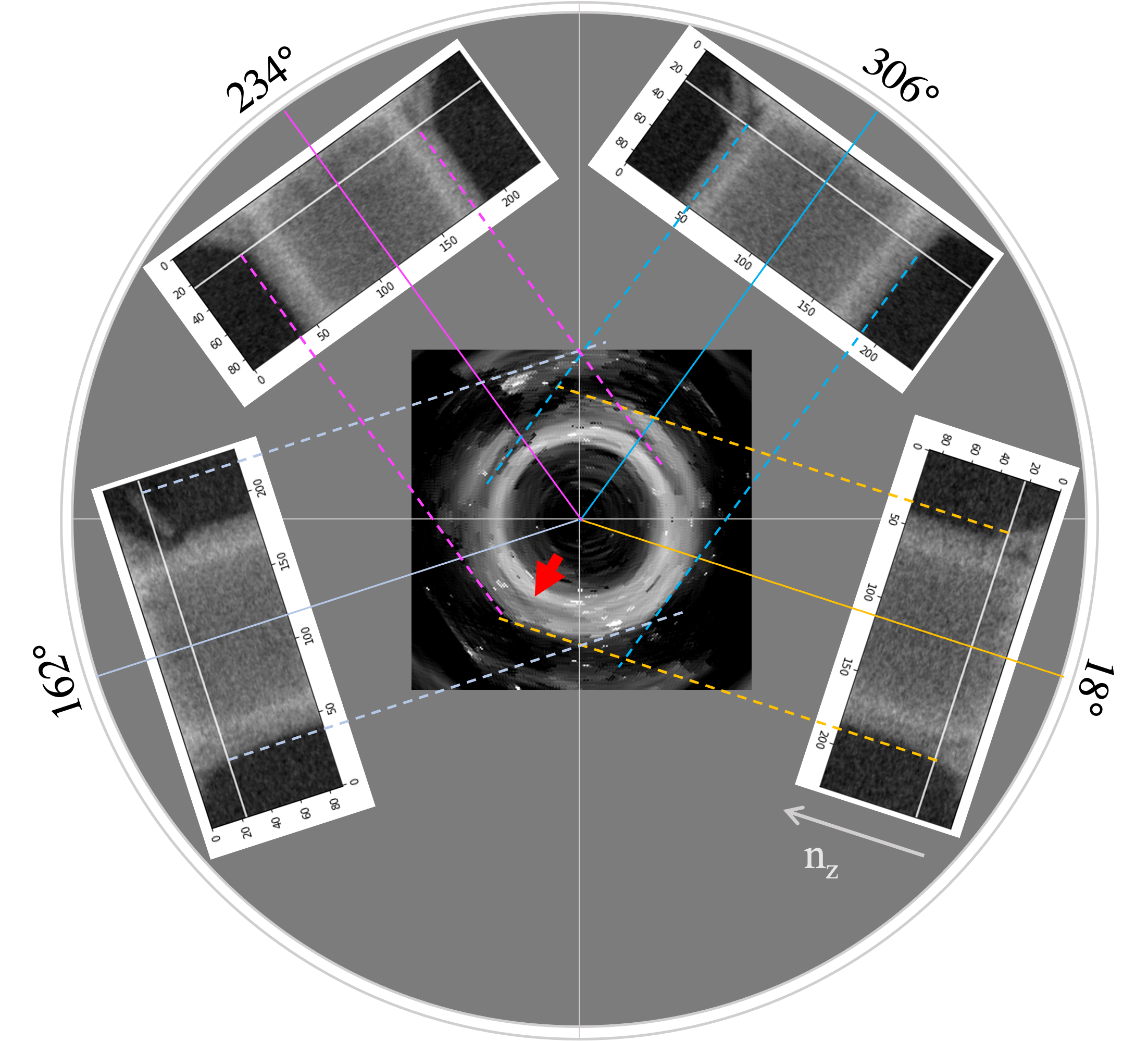}}
  \vspace{-0.1cm}
  \centerline{(g) MBIR/CTV}\medskip
\end{minipage}
\hfill
\vspace{-0.1cm}
\caption{Reconstructions of the experimentally imaged sandstone ring using various methods. The location of the cross section is marked by a white line in the radiographic measurements. The MBIR/CTV result in (g) is able to capture fine detail (see red arrow) that is consistent with features in radiographs at $18^{\circ}$ and $234^{\circ}$.}
\label{fig:real_results}
\end{figure*}

Table (\ref{table:sim_table}) compares PSNR values and run times for each method.
The CPU time is the total time required by the allocated resources,
while the wall time is the actual time elapsed.
The radial and transverse cracks aligned with view angles are reconstructed well
but the cracks oriented at $90^{\circ}$ are not preserved,
indicating that the MAP formulation is still underdetermined.
The concentric crack and material homogeneity are
captured best by the proposed method
while streak artifacts from the AWGN in the sinogram are largely suppressed.
Our previous approach, CE/MSF/BM3D/RA, tends to attenuate radial features far from center of the
image but these features are preserved by the proposed method. Note that the second-best performing method in terms of PSNR,
PPP/BM4D,is much more computationally expensive.

We also compare performance using a real dataset of a 3D printed sandstone ring imaged
at the moment of impact in the Kolsky experiment.
The hyperparameters for each result in~\fig{real_results}
were tuned to obtain the best contrast for the cracks visible in the radiographs.
MBIR/CTV is the best at suppressing streak artifacts and preserving
cracks in the reconstruction. As for the simulated dataset, MBIR/CTV is also among the best methods in terms of computation time.

\section{Conclusion}
\label{sec:conclusion}

The proposed approach represents a numerically stable and computationally efficient method of
computing the Total Variation penalty in cylindrical coordinates. Our experiments indicate that
it outperforms a number of other state-of-the-art priors in a problem involving ultra-sparse angle CT
of an object with approximate cylindrical symmetry, including a previous approach specifically
designed for the same type of problem~\cite{hossain2020ultrasparse}.

\bibliographystyle{IEEEtranD}
\bibliography{refs}

\begin{thebibliography}{10}
\providecommand{\url}[1]{#1}
\csname url@samestyle\endcsname
\providecommand{\newblock}{\relax}
\providecommand{\bibinfo}[2]{#2}
\providecommand{\BIBentrySTDinterwordspacing}{\spaceskip=0pt\relax}
\providecommand{\BIBentryALTinterwordstretchfactor}{4}
\providecommand{\BIBentryALTinterwordspacing}{\spaceskip=\fontdimen2\font plus
\BIBentryALTinterwordstretchfactor\fontdimen3\font minus
  \fontdimen4\font\relax}
\providecommand{\BIBforeignlanguage}[2]{{%
\expandafter\ifx\csname l@#1\endcsname\relax
\typeout{** WARNING: IEEEtran.bst: No hyphenation pattern has been}%
\typeout{** loaded for the language `#1'. Using the pattern for}%
\typeout{** the default language instead.}%
\else
\language=\csname l@#1\endcsname
\fi
#2}}
\providecommand{\BIBdecl}{\relax}
\BIBdecl

\bibitem{liao2018flash}
H.~Liao, ``Flash {X}-ray tomography of {K}olsky bar experiments,'' Ph.D.
  dissertation, Dept. Aero. and Astro. Eng., Purdue Univ., West Lafayette, IN,
  2018.

\bibitem{bouman1996unified}
C.~A. Bouman and K.~Sauer, ``A unified approach to statistical tomography using
  coordinate descent optimization,'' \emph{IEEE Transactions on Image
  Processing}, vol.~5, no.~3, pp. 480--492, 1996.

\bibitem{rudin1992nonlinear}
L.~I. Rudin, S.~Osher, and E.~Fatemi, ``Nonlinear total variation based noise
  removal algorithms,'' \emph{Physica D: Nonlinear Phenomena}, vol.~60, no.
  1-4, pp. 259--268, 1992.

\bibitem{thibault2007three}
J.-B. Thibault, K.~D. Sauer, C.~A. Bouman, and J.~Hsieh, ``A three-dimensional
  statistical approach to improved image quality for multislice helical {CT},''
  \emph{Medical Physics}, vol.~34, no.~11, pp. 4526--4544, 2007.

\bibitem{venkatakrishnan2013plug}
S.~V. Venkatakrishnan, C.~A. Bouman, and B.~Wohlberg, ``Plug-and-play priors
  for model based reconstruction,'' in \emph{IEEE Global Conference on Signal
  and Information Processing}, 2013, pp. 945--948.

\bibitem{kamilov-2023-plug}
U.~Kamilov, C.~A. Bouman, G.~T. Buzzard, and B.~Wohlberg, ``Plug-and-play
  methods for integrating physical and learned models in computational
  imaging,'' \emph{IEEE Signal Processing Magazine}, vol.~40, no.~1, pp.
  85--97, Jan. 2023.  \doi{10.1109/MSP.2022.3199595}

\bibitem{buzzard2018plug}
G.~T. Buzzard, S.~H. Chan, S.~Sreehari, and C.~A. Bouman, ``Plug-and-play
  unplugged: Optimization-free reconstruction using consensus equilibrium,''
  \emph{SIAM Journal on Imaging Sciences}, vol.~11, no.~3, pp. 2001--2020,
  2018.

\bibitem{dabov2007image}
K.~Dabov, A.~Foi, V.~Katkovnik, and K.~Egiazarian, ``Image denoising by sparse
  3-d transform-domain collaborative filtering,'' \emph{IEEE Transactions on
  Image Processing}, vol.~16, no.~8, pp. 2080--2095, 2007.

\bibitem{maggioni2012nonlocal}
M.~Maggioni, V.~Katkovnik, K.~Egiazarian, and A.~Foi, ``Nonlocal
  transform-domain filter for volumetric data denoising and reconstruction,''
  \emph{IEEE Transactions on Image Processing}, vol.~22, no.~1, pp. 119--133,
  2012.

\bibitem{majee20194d}
S.~Majee, T.~Balke, C.~A. Kemp, G.~T. Buzzard, and C.~A. Bouman, ``4d x-ray
  {CT} reconstruction using multi-slice fusion,'' in \emph{IEEE International
  Conference on Computational Photography (ICCP)}, 2019, pp. 1--8.

\bibitem{hossain2020ultrasparse}
M.~Hossain, S.~C. Paulson, H.~Liao, W.~W. Chen, and C.~A. Bouman,
  ``Ultra-sparse view reconstruction for flash {X}-ray imaging using consensus
  equilibrium,'' in \emph{54th Asilomar Conference on Signals, Systems, and
  Computers}, 2020, pp. 631--635.  \doi{10.1109/IEEECONF51394.2020.9443350}

\bibitem{bayram2012directional}
I.~Bayram and M.~E. Kamasak, ``A directional total variation,'' in
  \emph{Proceedings of the 20th European Signal Processing Conference
  (EUSIPCO)}, 2012, pp. 265--269.

\bibitem{zhang2013adaptive}
J.~Zhang, R.~Lai, and C.-C.~J. Kuo, ``Adaptive directional total-variation
  model for latent fingerprint segmentation,'' \emph{IEEE Transactions on
  Information Forensics and Security}, vol.~8, no.~8, pp. 1261--1273, 2013.

\bibitem{zhang2013edge}
H.~Zhang and Y.~Wang, ``Edge adaptive directional total variation,'' \emph{The
  Journal of Engineering}, vol. 2013, no.~11, pp. 61--62, 2013.

\bibitem{ehrhardt2016multicontrast}
M.~J. Ehrhardt and M.~M. Betcke, ``Multicontrast {MRI} reconstruction with
  structure-guided total variation,'' \emph{SIAM Journal on Imaging Sciences},
  vol.~9, no.~3, pp. 1084--1106, 2016.

\bibitem{asaki-2005-abel}
T.~J. Asaki, R.~Chartrand, K.~R. Vixie, and B.~Wohlberg, ``Abel inversion using
  total-variation regularization,'' \emph{Inverse Problems}, vol.~21, no.~6,
  pp. 1895--1903, 2005.  \doi{10.1088/0266-5611/21/6/006}

\bibitem{chen-2013-limitedangle}
Z.~Chen, X.~Jin, L.~Li, and G.~Wang, ``A limited-angle {CT} reconstruction
  method based on anisotropic {TV} minimization,'' \emph{Physics in Medicine
  and Biology}, vol.~58, no.~7, pp. 2119--2141, Mar. 2013.
  \doi{10.1088/0031-9155/58/7/2119}

\bibitem{guo2017image}
Y.~Guo, L.~Zeng, C.~Wang, and L.~Zhang, ``Image reconstruction model for the
  exterior problem of computed tomography based on weighted directional total
  variation,'' \emph{Applied Mathematical Modelling}, vol.~52, pp. 358--377,
  2017.

\bibitem{sheng2020sequential}
W.~Sheng, X.~Zhao, and M.~Li, ``A sequential regularization based image
  reconstruction method for limited-angle spectral {CT},'' \emph{Physics in
  Medicine \& Biology}, vol.~65, no.~23, p. 235038, 2020.

\bibitem{sijbers2004reduction}
J.~Sijbers and A.~Postnov, ``Reduction of ring artifacts in high resolution
  micro-{CT} reconstructions,'' \emph{Physics in Medicine \& Biology}, vol.~49,
  no.~14, p. N247, 2004.

\bibitem{lou-2015-weighted}
Y.~Lou, T.~Zeng, S.~Osher, and J.~Xin, ``A weighted difference of anisotropic
  and isotropic total variation model for image processing,'' \emph{SIAM
  Journal on Imaging Sciences}, vol.~8, no.~3, pp. 1798--1823, 2015.
  \doi{10.1137/14098435}

\bibitem{esser2010general}
E.~Esser, X.~Zhang, and T.~F. Chan, ``A general framework for a class of first
  order primal-dual algorithms for convex optimization in imaging science,''
  \emph{SIAM Journal on Imaging Sciences}, vol.~3, no.~4, pp. 1015--1046, 2010.

\bibitem{boyd2011distributed}
S.~Boyd, N.~Parikh, E.~Chu, B.~Peleato, J.~Eckstein \emph{et~al.},
  ``Distributed optimization and statistical learning via the alternating
  direction method of multipliers,'' \emph{Foundations and Trends in Machine
  learning}, vol.~3, no.~1, pp. 1--122, 2011.

\bibitem{sreehari-2016-plug}
S.~Sreehari, S.~V. Venkatakrishnan, B.~Wohlberg, G.~T. Buzzard, L.~F. Drummy,
  J.~P. Simmons, and C.~A. Bouman, ``Plug-and-play priors for bright field
  electron tomography and sparse interpolation,'' \emph{IEEE Transactions on
  Computational Imaging}, vol.~2, no.~4, pp. 408--423, Dec. 2016.
  \doi{10.1109/TCI.2016.2599778}

\bibitem{svmbir-2020}
{SVMBIR Development Team}, ``{S}uper-{V}oxel {M}odel {B}ased {I}terative
  {R}econstruction ({SVMBIR}),'' Software library available from
  \url{https://github.com/cabouman/svmbir}, 2020.

\bibitem{bm3d-2022}
Y.~M\"{a}kinen, ``Python wrapper for {BM3D} denoising,'' Software library
  available from \url{https://pypi.org/project/bm3d/}, 2022.

\bibitem{makinen-2019-exact}
Y.~M\"akinen, L.~Azzari, and A.~Foi, ``Exact transform-domain noise variance
  for collaborative filtering of stationary correlated noise,'' in \emph{IEEE
  International Conference on Image Processing (ICIP)}, Sep. 2019, pp.
  185--189.  \doi{10.1109/ICIP.2019.8802964}

\bibitem{bm4d-2022}
Y.~M\"{a}kinen, ``Python wrapper for {BM4D} denoising,'' Software library
  available from \url{https://pypi.org/project/bm4d/}, 2022.

\bibitem{balke-2022-scico}
T.~Balke, F.~Davis, C.~Garcia-Cardona, S.~Majee, M.~McCann, L.~Pfister, and
  B.~Wohlberg, ``Scientific computational imaging code {(SCICO)},''
  \emph{Journal of Open Source Software}, vol.~7, no.~78, p. 4722, Oct. 2022.
  \doi{10.21105/joss.04722}

\end{thebibliography}

\end{document}